\documentclass[twocolumn,preprintnumbers]{revtex4}
\usepackage{bm,enumerate,dcolumn,graphicx,amsmath,amssymb}
\newcommand{\comment}[1]{}

\begin{document}
\vspace{1.0 mm}
\title{Phase-dependent interference fringes in the wavelength scaling of harmonic efficiency}
\author{I. Yavuz$^{1,2}$*, E. A. Bleda$^3$, Z. Altun$^1$ and T. Topcu$^4$}
\address{$^1$Department of Physics, Marmara University, 34722,
Ziverbey, Istanbul, TURKEY.\\
$^2$Department of Chemistry and Biochemistry, University of California, Los Angeles, 90095 CA USA.\\
$^3$Department of Mathematics and Computer Science, Istanbul Arel University, 
Istanbul,TURKEY\\
$^4$Department of Physics, University of Nevada, Reno, NV 89557-0208 USA.\\
*corresponding author: ilhan.yavuz@marmara.edu.tr}

\begin{abstract}

We describe phase-dependent wavelength scaling of high-order harmonic generation efficiency driven by ultra-short laser fields in the mid-infrared. We employ both numerical solution of the time-dependent Schr\"{o}dinger equation and the Strong Field Approximation to analyze the fine-scale oscillations in the harmonic yield in the context of channel-closing effects. We show, by varying the carrier-envelope phase, that the amplitude of these oscillations depend strongly on the number of returning electron trajectories. Furthermore, the peak positions of the oscillations vary significantly as a function of the carrier-envelope phase. Owing to its practical applications, we also study the wavelength dependence of harmonic yield in the "single-cycle" limit, and observe a smooth variation in the wavelength scaling originating from the vanishing fine-scale oscillations. 

\end{abstract}

\maketitle
\section{Introduction}

High-order harmonic generation (HHG), has attracted immense interest due to its novel applications in ultra-fast science~\cite{1p,1,2,3}. From a semi-classical viewpoint, the electron dynamics leading to the HHG process can simply be described by the three-step model \cite{4}, in which (1) an electron is detached from an atom by a laser field; (2) it propagates in the continuum until (3) its recombination with the atom, accompanied by photon emission. The maximum photon energy that can be generated is given by the cut-off law 
${{\omega }_{c}}=\left| {{E}_{b}} \right|+3.17{{U}_{p}}$, 
where $\left| {{E}_{b}} \right|$ is the binding energy of the active electron \cite{4}. $3.17{{U}_{p}}$ is the maximum kinetic energy acquired in the laser field and ${{U}_{p}}$ is the ponderomotive potential: 
${{U}_{p}}\sim {\ }I{{\lambda }^{2}}$.

In recent years, there has been renewed interest in the fundamental aspects of HHG, primarily based on the scaling of its yield in terms of both the laser wavelength and intensity~\cite{6}. It has been shown that the integrated harmonic yield, obtained by integrating the power spectrum $S(\omega)$ within a fixed photon energy interval, {\it i.e.}, 
\begin{eqnarray}\label{eq:Delta_S}
\Delta S=\int_{{{\omega }_\text{i}}}^{{{\omega }_\text{f}}}{S(\omega )d\omega } \;,
\end{eqnarray}
contrasts with the conventional ${{\lambda}^{-3}}$ scaling of the harmonic yield, predicted as early as in the mid-90s~\cite{5}, which originates from wave-packet spread. Recent theoretical calculations of single atom response and also experimental observations (designed so that macroscopic effects are minimized) revealed a rapid decline in harmonic efficiency, scaling as $\sim{{\lambda}^{-5}}$-${{\lambda}^{-6}}$, suggesting that global scaling can depend on the field intensity and atomic species of interest ~\cite{6,7,8,8bucuk,9,9bucuk1,9bucuk}. 

It has recently been proposed that the variations in the overall wavelength scaling can also be phase-dependent in the limit of ultra-short laser fields ~\cite{10}. The scaling remains roughly $\sim{{\lambda}^{-5}}$, but a substantial variation with carrier-envelope phase emerges. For molecules, on the other hand, a more dramatic decrease is predicted. This is attributed to fast ground-state depletion forced by interatomic (vibrational) motion excited by radiation at longer wavelengths ~\cite{11,12}. 

There are a number of studies addressing the origin of this rapid decrease in harmonic yield as well as its intensity and target atom dependence as a function of the laser wavelength ~\cite{6,7,8,8bucuk,9,9bucuk1,9bucuk}. According to ~\cite{7,9bucuk1}, the $I{{\lambda}^{2}}$ scaling of the size of the plateau region in the harmonic spectrum in a classical framework can bring an additional $\sim{{\lambda}^{-2}}$ factor in the overall scaling. Along with the $\sim{{\lambda}^{-3}}$ scaling due to the wave-packet spread, this results in $\sim{{\lambda}^{-5}}$ for overall scaling, consistent with quantum mechanical predictions ~\cite{6,7,8}.  However, a factor of $\sim{{\lambda}^{-2}}$ originates from the definition of the integrated harmonic yield~\eqref{eq:Delta_S} itself since  $d\omega \propto {{\lambda }^{-2}}d\lambda $. 

Schiessl {\it et al.}~\cite{7} have shown that in the 800-2000 nm wavelength range, harmonic yield exhibits a series of rapid fluctuations with a period of 6-20 nm instead of a smooth decrease. They have identified the origin of these fine-scale oscillations as the interference fringes of rescattering electron trajectories. Frolov {\it et al.}~\cite{8} have shown that the peak positions of these fringes correspond to enhancement of harmonic yield due to opening of a multiphoton ionization channel. The number of photons $R$ the atom needs to absorb to escape through this channel is called the ``channel closing number", and obeys~\cite{13,14} 
\begin{eqnarray}\label{eq:photon_num}
R{{\omega }_{0}}-({{I}_{p}}+{{U}_{p}})=0 \;,
\end{eqnarray}
where $\omega_0$ is the laser frequency and $I_p$ is the ionization potential of the target atom. 

In a typical HHG process, the ionization dynamics falls into the tunneling regime (characterized by the Keldysh parameter, $\gamma<1$). However, it has been recently demonstrated~\cite{TopRob12} that $\gamma<1$ is not a sufficient condition by itself for tunneling to dominate. If the minimum number of photons that needs to be absorbed for electron escape is small enough, the ionization can be dominated by photo-absorption even when $\gamma <1$. In this work, the Keldysh parameter ranges between $\sim$0.40 and $\sim$0.50, and the atom is $\sim$40 photons below the ionization threshold, which is why these multiphoton ionization induced oscillations are referred to 
as ``fine-scale" oscillations. 

The channel closing number in Eq.~\eqref{eq:photon_num} can also be expressed in terms of the laser wavelength $\lambda$
\begin{eqnarray}\label{eq:photon_num_wlen}
R = \frac{\lambda}{2\pi c} \left( I_p + U_p(\lambda) \right) \;.
\end{eqnarray}
As $\lambda$ is varied, the atom falls in and out of resonance with multiphoton ionization channels where the minimum number of photons to absorb for ionization is $R$. 

In this report, we investigate the phase-dependence of the fine-scale oscillations observed in the HHG yield in terms of the channel-closing number $R$. In addition to the overall phase-dependent scaling, the amplitude of these oscillations also varies with the phase of the laser field. We find that the variations in oscillations can be compensated by a specific choice of the carrier-envelope phase (CEP). We contrast our results with those obtained in the ``single-cycle" limit. We use atomic units are used throughout, unless specified otherwise. 

\section{Methodology}
We start by describing our numerical calculations. We study the response of a single atom to a linearly-polarized strong laser field based on the numerical solution of the time-dependent Schr\"{o}dinger equation (TDSE) and semi-classical Strong Field Approximation (SFA)~\cite{5} in the length gauge. 

The TDSE is expressed as 
\begin{eqnarray}\label{eq:tdse}
i\frac{\partial \psi ({\bf{r}},t)}{\partial t}=\left[ -\frac{1}{2}{{\nabla }^{2}}+V(r)+zF(t) \right]\psi ({\bf{r}},t) \;.
\end{eqnarray}
The target atom is Hydrogen, therefore $V(r)=-1/r$. The electric field of the laser pulse is $F(t)={F_0}\exp [-(4\ln 2){{t}^{2}}/{{\tau }^{2}}]\cos ({{\omega }_{0}}t+{{\phi }_{CEP}})$, where $F_0$ is the peak field strength and $\tau $ is the field duration at FWHM. We use a two-cycle pulse at FWHM in our calculations. The numerical solution of the TDSE is carried out by a split-operator technique~\cite{14bucukx}, where the time-dependent wave-function is of the form 
$\psi ({\bf{r}},t)=\sum\nolimits_{l}{{{R}_{l}}(r,t)Y_{l}^{0}({\mathbf{\hat{r}}})/r}$. 
In order to avoid spurious reflections from grid boundaries, a smooth cut-off function starting from 2/3 of the radial box is multiplied by the solutions of TDSE at each time step. 

In order to evaluate the time dependent dipole in SFA, we assume that the ground state wave-function has the form $\psi (r)=\exp [-r^2/2{{\delta }^{2}}]/\left( \pi \delta  \right)^{3/4}$~\cite{5}, where $\delta$ is a tunable parameter to fit the $I_p$ under study. Using this formula one can  calculate the dipole moment analytically~\cite{5,14bucuk2}: 
\begin{equation}\label{eq:sfa_dip}
\begin{split}
 d(t)=&\frac{i}{{{\delta }^{7}}}\int_{-\infty }^{t}{d{t}'{{[2C(t,{t}')]}^{3/2}}F({t}')\{A(t)A({t}')+C(t,{t}')}\\
 & \{1-D(t,{t}')[A(t)+A({t}')]\}+{{C}^{2}}(t,{t}'){{D}^{2}}(t,{t}')\} \\ 
 & \times \exp \left( -i\{{{I}_{p}}\tau+B(t,{t}')\} \right. \\ 
 & \left. -\frac{{{\delta }^{2}}[{{A}^{2}}(t)+{{A}^{2}}({t}')]-C(t,{t}'){{D}^{2}}(t,{t}')}{2} \right) \;. 
\end{split}
\end{equation}
The time-dependent functions $B(t,{t}')$, $C(t,{t}')$ and $D(t,{t}')$ in this expression are given by
\begin{align}
  B(t,{t}')=&\frac{1}{2}\int_{t}^{{{t}'}}{d{t}''{{A}^{2}}({t}'')} \\ 
  C(t,{t}')=&[2{{\delta }^{2}}+i\tau ] \\ 
  D(t,t)=&[A(t)+A({t}')]{{\delta }^{2}}+i\int_{{{t}'}}^{t}{d{t}''A({t}'')} 
\end{align}
where $\tau = t-{t}'$ and $A(t)=-\int{F(t)dt}$ is the vector potential. We calculate harmonic spectrum by the Fourier transforming the dipole moment $d(t)$ in our SFA calculations and dipole acceleration~\cite{14bucuk1} 
\begin{eqnarray}\label{eq:tdse_dacc}
a(t)=\int \psi^{*} ({\bf{r}},t)[-\frac{\partial}{\partial z}V(r) ] \psi ({\bf{r}},t) d^3{\bf{r}}
\end{eqnarray}
in our TDSE calculations. 

The integrated harmonic yield $\Delta S$, on the other hand, is calculated by integrating the
radiated power $S(\omega)$ over a fixed energy interval $\Delta E = \omega_\text{f} - \omega_\text{i}$ using Eq.~\eqref{eq:Delta_S}~\cite{6,7} where 
\begin{eqnarray}\label{eq:gauges}
 S(\omega) &=& \frac{2}{{3\pi c^3}} {{\left| a(\omega ) \right|}^{2}} {\text{ (acceleration gauge)}} \;, \\
 S(\omega) &=& \frac{2 \omega ^{4}}{{3\pi c^3}} {{\left| d(\omega ) \right|}^{2}} {\text{ (length gauge)}} \;.
\end{eqnarray}
Here $a(\omega)$ and $d(\omega)$ are the Fourier transforms of the dipole acceleration and the dipole moment, respectively. 

\begin{figure}[t]
  \begin{center}
        \resizebox{85mm}{!}{\includegraphics{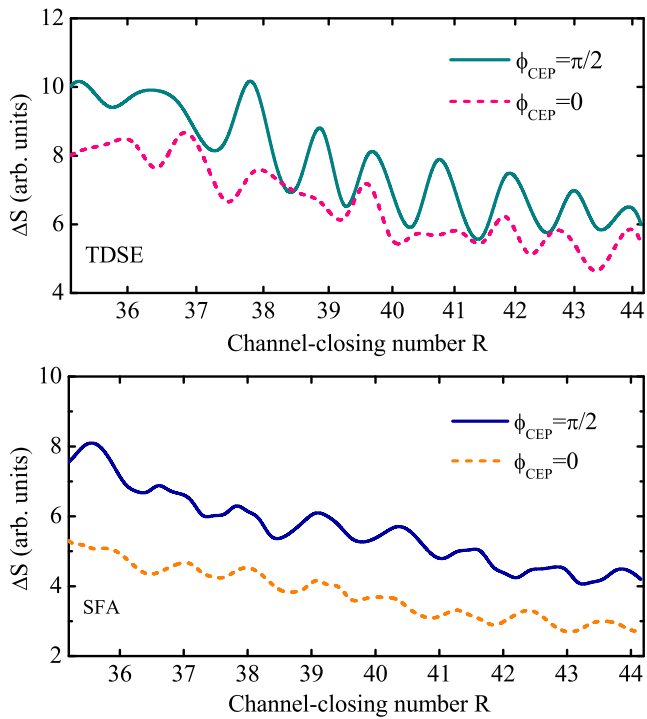}}
  \end{center}
  \caption{ (Color online) TDSE (upper panel) and SFA (lower panel) calculations for $\Delta S$ as a function of the channel-closing number $R$ for two different values of $\phi_{CEP}$ using a 2-cycle pulse at 320 TW/cm$^2$. In the evaluation of $\Delta S$, we integrated $S(\omega)$ over the energy interval of $45-75$ eV. 
  }
  \label{fig_1}
\end{figure}

\section{Fine-scale oscillations and wavelength scaling in HHG yield}
Instead of presenting the integrated yield $\Delta S$ as function of the laser wavelength, we present our results in the channel-closing number $R$ (longer $\lambda$ means larger $R$). As $\lambda$ is varied, the atom falls in and out of resonance with multiphoton ionization channels while still in the tunneling regime. For longer $\lambda$, it takes more photons to reach the ionization threshold and $R$ increases. Fig.~\ref{fig_1} presents our results for $\Delta S$, as a function of the channel-closing number $R$ for two different CEP values calculated by solving the TDSE and using the SFA model. The $R$ values are between 35 and 44, which fall into the wavelength range 1000-1100 nm at 320 TW/cm$^2$.  Both curves show a distinct motif, where in the case of $\phi_{CEP}$ $=\pi /2$, the oscillations are clearly visible in agreement with the previous studies~\cite{7,8,9bucuk1}. However, they are less clear for $\phi_{CEP}$ $=0$ and shifted towards lower $R$ relative to those for $\phi_{CEP}$ $=\pi /2$. 

An interesting feature in Fig.~\ref{fig_1} is that the peak positions in Fig.~\ref{fig_1} do not always correspond to integer values of $R$. However, we find that the peaks are essentially equally spaced. This can be understood in terms of the AC Stark shift  experienced by a real atom in a strong laser field. Because the field is not perturbative, the shift cannot simply be expressed using second order perturbation theory. An alternative way of looking at this shift is as a depression of the ionization threshold by the ponderomotive potential. Because $U_p$ is the quiver energy of a free electron in a laser field, it does not include information about the atom itself. The structure information about the atom is very simply encoded in $I_p$ in Eq.~\eqref{eq:photon_num}. Effectively, this is saying that regardless of atomic structure, the ground state is always shifted  by the same amount $U_p$. This approximation introduces ambiguity in the ground state energy shift, and this ambiguity is typically blamed on $I_p$ entirely. In~\cite{8,9bucuk1}, various models for effective ionization potentials are used that better match the peak positions with integer values of $R$. 

\begin{figure}[ht]
  \begin{center}
    \resizebox{85mm}{!}{\includegraphics{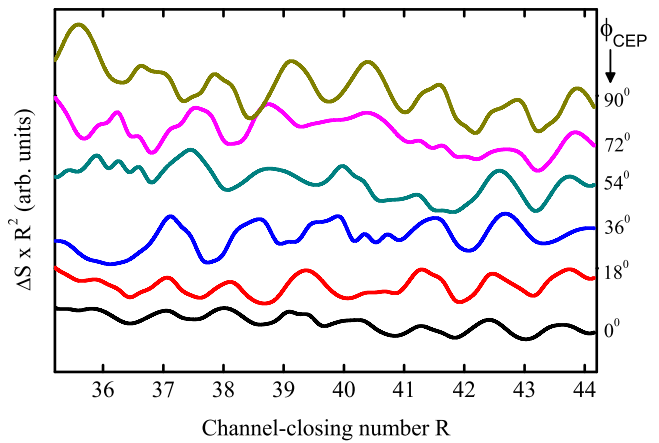}}
  \end{center}
  \caption{ (Color online) Based on the SFA model, fine-scale oscillations in the wavelength scaling as a function of $R$ for 6 different $\phi_{CEP}$ between 0 and $\pi/2$. For each value of $\phi_{CEP}$, $\Delta S$ is multiplied by $R^2$ to reduce the overall decrease with $R$ and the curves are vertically shifted for clarity.
  }
  \label{fig_sfa_phase}
\end{figure}

The weakening of the peaks going from $\phi_{CEP}=\pi/2$ to 0 in Fig.~\ref{fig_1} can be explained by the interference of multiple returning quantum trajectories~\cite{7,9bucuk1}. It has been demonstrated that the increase in the number of quantum trajectories leads to fine-scale oscillations in a multi-cycle laser field~\cite{7}.  In the case of few-cycle pulses, the number of returning quantum trajectories are smaller and are CEP dependent. In Fig.~\ref{fig_1}, the number of cycles at FWHM are less for the $\phi_{CEP}$ $=0$ pulse than for the $\phi_{CEP}$ $=\pi /2$ pulse, leading to smaller number of interfering electron trajectories than for $\phi_{CEP}$ $=\pi /2$. This weakens the amplitudes of the oscillations in  Fig.~\ref{fig_1} in agreement with previous studies~\cite{6,7,8}. 
Also clear from Fig.~\ref{fig_1} is that the amplitudes of the oscillations do not depend on the field strength since the maximum field amplitude for the electric field is only $\sim$4\% lower for $\phi_{CEP}$ $=\pi /2$ than that for $\phi_{CEP}$ $=0$.

\begin{figure}[ht]
  \begin{center}
    \resizebox{85mm}{!}{\includegraphics{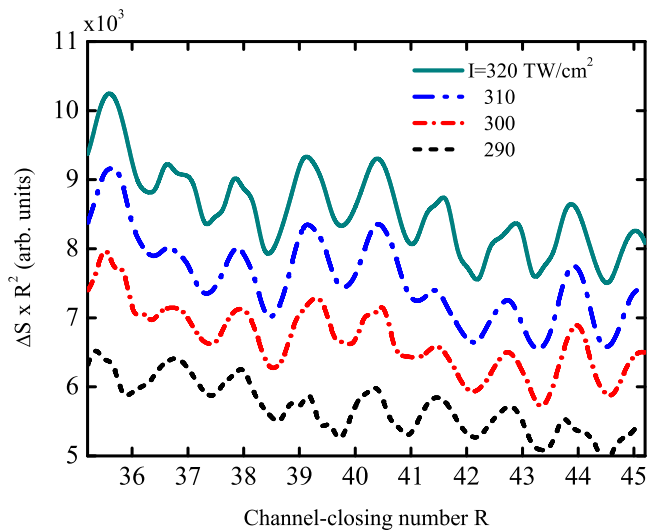}}
  \end{center}
  \caption{ (Color online) Variation of fine-scale oscillations in wavelength scaling as a function of $R$ for
different intensities in the SFA model. In each plot, $\phi_{CEP}=\pi/2$ and $\Delta S$ are multiplied by $R^2$ to reduce the overall decrease with $R$. 
  }
  \label{fig_sfa_intensity}
\end{figure}

The effect of CEP on the structure of the oscillations is less pronounced from the SFA model in the lower panel of Fig.~\ref{fig_1}. The differences between the TDSE and the SFA model can be attributed to the approximations made in SFA, {\it e.g.} neglecting the excited states of the target atom and the Coulomb-tail when electron is in the continuum. Apart from diminishing in the oscillation amplitudes, the bottom panel of Fig.~\ref{fig_1} displays the same features seen from the TDSE calculations with regard to the CEP dependence and the peak positions. Variation of the peak positions according to the SFA model for a series of CEP values between those seen in Fig.~\ref{fig_1} can be seen in Fig.~\ref{fig_sfa_phase}. Again, some the peaks are slightly shifted from integer $R$ as $\phi_{CEP}$ is varied, and their amplitudes become smaller as $\phi_{CEP}$ is reduced. The peaks are separated by essentially one photon: {\it e.g.} in the $41\le R\le 44$ range, $\left\langle \delta R \right\rangle \approx 1.1$, which translates into $\left\langle \delta \lambda \right\rangle \approx 12\text{ }nm$. Here $\left\langle..\right\rangle$ denotes CEP average over the $0\le {{\phi }_{CEP}}\le \pi /2$ interval. 

Another relavant laser parameter of interest that can potentially affect these oscillations is intensity. The intensity dependence of the fine-scale oscillations in $\Delta S$ are plotted in Fig.~\ref{fig_sfa_intensity} for a two-cycle laser pulse, where intensities vary between 290 and 320 TW/cm$^2$ with $\phi_{CEP}=\pi/2$. In this case, we observe no dramatic differences in the oscillation patterns for various field intensities except that their overall efficiencies increase with increasing intensity. Furthermore, the amplitudes of the oscillations slightly increase with intensity. The peak positions remain the same as intensity is varied meaning that the modulation period $\delta \lambda$ is independent of field intensity we consider. This is mainly because the peak electric field strength changes only by $\sim$5\% from the lowest to the highest intensity in Fig.~\ref{fig_sfa_intensity}.

\begin{figure}[t]
  \begin{center}
      \resizebox{85mm}{!}{\includegraphics{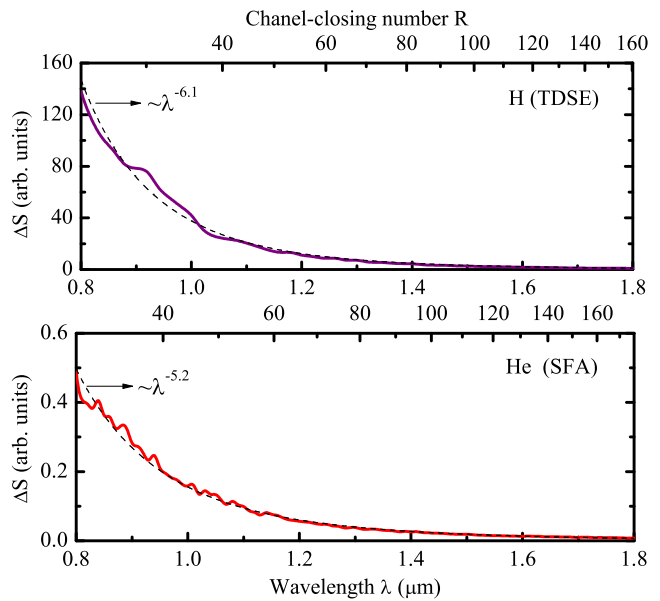}}
  \end{center}
  \caption{ (Color online) Variation of $\Delta S$ as a function of $\lambda$ in the single-cycle limit. In calculating $\Delta S$, $S(\omega)$ is integrated over photon energies between 50-70 eV for $I=320$ TW/cm$^2$. We use the TDSE model for H, whereas for He the SFA model is used. The conventional power laws $\lambda^{-x}$ are also indicated in the figures. $\gamma$ ranges from 0.60 to 0.27 for H, and from 0.80 to 0.36 for He. 
  }
  \label{fig_atom_specific} 
\end{figure}

So far we looked at wavelength dependence of $\Delta S$ for atoms driven by a two-cycle pulse. Wavelength scaling of HHG yield in the "single-cycle" limit is also of great interest, since HHG is being used as a way of generating coherent attosecond pulse trains as well as single attosecond pulses~\cite{15,16}. The main difference between the multi-cycle and single-cycle pulses is that the former support multiple electron return trajectories and attosecond pulse trains. For single-cycle pulses with a specific value of CEP, single return trajectory, and as a result, single harmonic emission is possible, which is desired for isolated attosecond pulse generation. 

It is clear from Eq.~\eqref{eq:photon_num} that the $R$ depends on the ionization potential of the target atom, meaning that the variations in $\Delta S \sim{{\lambda}^{-x}}$ scaling are atom specific. We performed calculations for the wavelength-scaling of $\Delta S$ by single-cycle pulses with $\phi_{CEP}=0$, where we chose H and He atoms as targets. For the H atom, we employ TDSE whereas for He the SFA model is used. To calculate $\Delta S$ at each value of the laser wavelength $\lambda$, we integrate over the spectral interval 50-70 eV for $I=320$ TW/cm$^{2}$. Fig.~\ref{fig_atom_specific} shows these results. $\Delta S$ shows an overall $\lambda^{-x}$ dependence with $x\simeq 5-6$. 
Since the fine-scale oscillations result from cycle-by-cycle interference of the returning electron trajectories, the amplitudes of these fluctuations are commensurate with the number of returning electron trajectories~\cite{7,9bucuk1,10}. This is the reason why these fluctuations do not appear in the single-cycle limit in Fig.~\ref{fig_atom_specific}. 

\section{Conclusions}

In this report, we studied fine-scale oscillations in the wavelength scaling of harmonic efficiency using few-cycle pulses where the effect of CEP is pronounced. Our simulations are based on the numerical solutions of TDSE and SFA model. We found that the fine-scale oscillations in wavelength dependence is CEP dependent and therefore can be controlled by phase-stabilization. Next, we employed both TDSE and SFA methods to characterize the $\lambda$-dependence of the HHG yield in single-cycle limit. We found that the fluctuations in the HHG yield vanish due to the absence of cycle-by-cycle interference for single-cycle pulses. 

\section{Acknowledgments}

IY, EAB and ZA are grateful to TUBITAK ULAKBIM, High Performance and Grid Computing Center, BAPKO of 
Marmara University. TT was supported by the National Science Foundation Grant No. PHY-1212482.

\end{document}